\documentclass[10pt,journal,compsoc]{IEEEtran}
\ifCLASSOPTIONcompsoc
  \usepackage[nocompress]{cite}
\else
  \usepackage{cite}
\fi
\ifCLASSINFOpdf
\else
\fi
\usepackage{booktabs} % For formal tables
\usepackage{adjustbox}
\usepackage{multirow}
\usepackage{amsmath}
\usepackage[linesnumbered,ruled]{algorithm2e}
\usepackage{caption}
\renewcommand\baselinestretch{0.97}
\begin{document}
%
% paper title
% Titles are generally capitalized except for words such as a, an, and, as,
% at, but, by, for, in, nor, of, on, or, the, to and up, which are usually
% not capitalized unless they are the first or last word of the title.
% Linebreaks \\ can be used within to get better formatting as desired.
% Do not put math or special symbols in the title.
\title{JointDNN: An Efficient Training and Inference Engine for Intelligent Mobile Cloud Computing Services}
%
%
% author names and IEEE memberships
% note positions of commas and nonbreaking spaces ( ~ ) LaTeX will not break
% a structure at a ~ so this keeps an author's name from being broken across
% two lines.
% use \thanks{} to gain access to the first footnote area
% a separate \thanks must be used for each paragraph as LaTeX2e's \thanks
% was not built to handle multiple paragraphs
%
%
%\IEEEcompsocitemizethanks is a special \thanks that produces the bulleted
% lists the Computer Society journals use for "first footnote" author
% affiliations. Use \IEEEcompsocthanksitem which works much like \item
% for each affiliation group. When not in compsoc mode,
% \IEEEcompsocitemizethanks becomes like \thanks and
% \IEEEcompsocthanksitem becomes a line break with idention. This
% facilitates dual compilation, although admittedly the differences in the
% desired content of \author between the different types of papers makes a
% one-size-fits-all approach a daunting prospect. For instance, compsoc 
% journal papers have the author affiliations above the "Manuscript
% received ..."  text while in non-compsoc journals this is reversed. Sigh.

\author{Amir Erfan Eshratifar,
        Mohammad Saeed Abrishami,
        and~Massoud Pedram,~\IEEEmembership{Fellow,~IEEE}
        \IEEEcompsocitemizethanks{\IEEEcompsocthanksitem All authors are with the Department of Electrical Engineering, University of Southern California, Los Angeles, CA 90089-2562 USA (eshratif@usc.edu, abri442@usc.edu, pedram@usc.edu).\protect\\}
}

\IEEEtitleabstractindextext{%

\begin{abstract}
Deep learning models are being deployed in many mobile intelligent applications. End-side services, such as intelligent personal assistants, autonomous cars, and smart home services often employ either simple local models on the mobile or complex remote models on the cloud. However, recent studies have shown that partitioning the DNN computations between the mobile and cloud can increase the latency and energy efficiencies. In this paper, we propose an efficient, adaptive, and practical engine, JointDNN, for collaborative computation between a mobile device and cloud for DNNs in both inference and training phase. JointDNN not only provides an energy and performance efficient method of querying DNNs for the mobile side but also benefits the cloud server by reducing the amount of its workload and communications compared to the cloud-only approach. Given the DNN architecture, we investigate the efficiency of processing some layers on the mobile device and some layers on the cloud server. We provide optimization formulations at layer granularity for forward- and backward-propagations in DNNs, which can adapt to mobile battery limitations and cloud server load constraints and quality of service. JointDNN achieves up to 18 and 32 times reductions on the latency and mobile energy consumption of querying DNNs compared to the status-quo approaches, respectively.
\end{abstract}

% Note that keywords are not normally used for peerreview papers.
}

% make the title area
\maketitle

\IEEEdisplaynontitleabstractindextext

\IEEEpeerreviewmaketitle

\ifCLASSOPTIONcompsoc
\section{Introduction}
DNN architectures are promising solutions in achieving remarkable results in a wide range of machine learning applications, including, but not limited to computer vision, speech recognition, language modeling, and autonomous cars. Currently, there is a major growing trend in introducing more advanced DNN architectures and employing them in end-user applications. The considerable improvements in DNNs are usually achieved by increasing computational complexity which requires more resources for both training and inference~\cite{Pouyanfar:2018:SDL:3271482.3234150}. Recent research directions to make this progress sustainable are: development of Graphical Processing Units (GPUs) as the vital hardware component of both servers and mobile devices~\cite{GPU4NN}, design of efficient algorithms for large-scale distributed training~\cite{DistributedDNN} and efficient inference~\cite{SAMRAGH}, compression and approximation of models~\cite{efficientDNN}, and most recently introducing collaborative computation of cloud and fog as known as dew computing~\cite{DewComputing}. 

Deployment of cloud servers for computation and storage is becoming extensively favorable due to technical advancements and improved accessibility. Scalability, low cost, and satisfactory Quality of Service (QoS) made offloading to cloud a typical choice for computing-intensive tasks. On the other side, mobile-device are being equipped with more powerful general-purpose CPUs and GPUs. Very recently there is a new trend in hardware companies to design dedicated chips to better tackle machine-learning tasks. For example, Apple's A11 Bionic chip~\cite{iPhoneX} used in iPhone X uses a neural engine in its GPU to speed up the DNN queries of applications such as face identification and facial motion capture~\cite{animoji}.

In the status-quo approaches, there are two methods for DNN inference: mobile-only and cloud-only. In simple models, a mobile device is sufficient for performing all the computations. In the case of complex models, the raw input data (image, video stream, voice, etc.) is uploaded to and then the required computations are performed on the cloud server. The results of the task are later downloaded to the device. The effects of raw input and feature compression are studied in \cite{bottlenet} and \cite{towards}.

Despite the recent improvements of the mobile devices mentioned earlier, the computational power of mobile devices is still significantly weaker than the cloud ones. Therefore, the mobile-only approach can cause large inference latency and failure in meeting QoS. Moreover, embedded devices undergo major energy consumption constraints due to battery limits. On the other hand, cloud-only suffers communication overhead for uploading the raw data and downloading the outputs. Moreover, slowdowns caused by service congestion, subscription costs, and network dependency should be considered as downsides of this approach~\cite{Eshratifar:2018:EPE:3194554.3194565}. 

The superiority and persistent improvement of DNNs depend heavily on providing a huge amount of training data. Typically, this data is collected from different resources and later fed into a network for training. The final model can then be delivered to different devices for inference functions. However, there is a trend of applications requiring adaptive learning in online environments, such as self-driving cars and security drones \cite{AdaptiveCar}\cite{EASI}. Model parameters in these smart devices are constantly being changed based on their continuous interaction with their environment. The complexity of these architectures with an increasing number of parameters and current cloud-only methods for DNN training implies a constant communication cost and the burden of increased energy consumption for mobile devices. The main difference of collaborative training and cloud-only training is that the data transferred in the cloud-only approach is the input data and model parameters but in the collaborative approach, it is layer(s)'s output and a portion of model parameters. Therefore, the amount of data communicated can be potentially decreased\cite{Neurosurgeon}.

Automatic partitioning of computationally extensive tasks over the cloud for optimization of performance and energy consumption has been already well-studied~\cite{CloneCloud}. Most recently, scalable distributed hierarchy structures between the end-user device, edge, and cloud have been suggested~\cite{CloudEdgeDevice} which are specialized for DNN applications. However, exploiting the layer granularity of DNN architectures for run-time partitioning has not been studied thoroughly yet.

\begin{figure}[h]
\centering
\includegraphics[width=\linewidth]{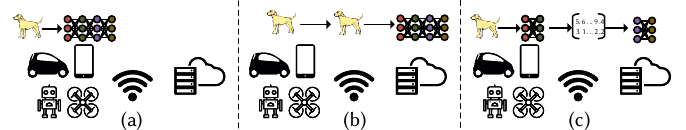}
\caption{Different computation partitioning methods. (a) Mobile only: computation is completely done on the mobile device. (b) Cloud only: raw input data is sent to the cloud server, computations is done on the cloud server and results are sent back to the mobile device. (c) JointDNN: DNN architecture is partitioned at the granularity of layers, each layer can be computed either on cloud or mobile.}
\captionsetup{justification=centering,margin=2cm}
\label{fig:fig_intro}
\end{figure}

% Story of the paper at the end of the introduction:
% Considering these char. we modeled it with a DAG
% Parameters of a DAG are
% Now you have a complete model, what do u want to solve, how do u want to solve it? Algorithm
% Parameters are extracted by experiments
% Results
% Conclusion
% Note: I did not mention the ILP for constraints! 
In this work, we are investigating the inference and training of DNNs in a \textbf{joint} platform of mobile and cloud as an alternative to the current single-platform methods as illustrated in Figure~\ref{fig:fig_intro}. Considering DNN architectures as an ordered sequence of layers, and the possibility of computation of every layer either on mobile or cloud, we can model the DNN structure as a Directed Acyclic Graph (DAG). The parameters of our real-time adaptive model are dependent on the following factors: mobile/cloud hardware and software resources, battery capacity, network specifications, and QoS. Based on this modeling, we show that the problem of finding the optimal computation schedule for different scenarios, i.e. best performance or energy consumption, can be reduced to the polynomial-time shortest path problem. 

To present realistic results, we made experiments with fair representative hardware of mobile device and cloud. To model the communication costs between platforms, we used various mobile network technologies and the most recent reports on their specifications in the U.S. 

DNN architectures can be categorized based on functionality. These differences enforce specific type and order of layers in architecture, directly affecting the partitioning result in the collaborative method. For discriminative models, used in recognition applications, the layer size gradually decreases going from input toward output as shown in Figure \ref{fig:DNN_category}. This sequence suggests the computation of the first few layers on the mobile device to avoid excessive communication cost of uploading large raw input data. On the other hand, the growth of the layer output size from input to output in generative models which are used for synthesizing new data, implies the possibility of uploading a small input vector to the cloud and later downloading one of the last layers and performing the rest of computations on the mobile device for better efficiency. Interesting mobile applications like image-to-image translation are implemented with autoencoder architectures whose middle layers sizes are smaller compared to their input and output. Consequently, to avoid huge communication costs, we expect the first and last layers to be computed on the mobile device in our collaborative approach. We examined eight well-known DNN benchmarks selected from these categories to illustrate their differences in the collaborative computation approach. 

As we will see in Section~\ref{Results}, the communication between the mobile and cloud is the main bottleneck for both performance and energy in the collaborative approach. We investigated the specific characteristics of CNN layer outputs and introduced a loss-less compression approach to reduce the communication costs while preserving the model accuracy. 

State-of-the-art work for collaborative computation of DNNs~\cite{Neurosurgeon} only considers one offloading point, assigning computation of its previous layers and next layers on the mobile and cloud platforms, respectively. We show that this approach is non-generic and fails to be optimal, and introduced a new method granting the possibility of computation on either platform for each layer independent of other layers. Our evaluations show that JointDNN significantly improves the latency and energy up to $3\times$ and $7\times$ respectively compared to the status-quo single platform approaches without any compression. The main contributions of this paper can be listed as:

\begin{itemize}
\item Introducing a new approach for the collaborative computation of DNNs between the mobile and cloud 
\item Formulating the problem of optimal computation scheduling of DNNs at layer granularity in the mobile cloud computing environment as the shortest path problem and integer linear programming (ILP) 
\item Examining the effect of compression on the outputs of DNN layers to improve communication costs
\item Demonstrating the significant improvements in performance, mobile energy consumption, and cloud workload achieved by using \textbf{JointDNN}
\end{itemize}

\begin{figure}
\centering
\includegraphics{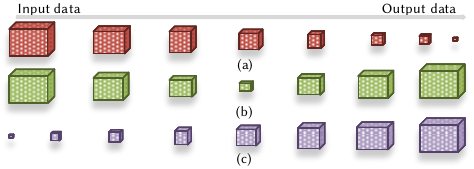}
\caption{Typical layer size in (a) Discriminative (b) Autoencoder (c) Generative models.}
\captionsetup{justification=centering,margin=2cm}
\label{fig:DNN_category}
\end{figure}

\else
\section{Introduction}
\label{sec:introduction}
\fi
% Computer Society journal (but not conference!) papers do something unusual
% with the very first section heading (almost always called "Introduction").
% They place it ABOVE the main text! IEEEtran.cls does not automatically do
% this for you, but you can achieve this effect with the provided
% \IEEEraisesectionheading{} command. Note the need to keep any \label that
% is to refer to the section immediately after \section in the above as
% \IEEEraisesectionheading puts \section within a raised box.
\section{Problem definition and modeling}
In this section, we explain the general architecture of DNN layers and our profiling method. Moreover, we elaborate on how cost optimization can be reduced to the shortest path problem by introducing the JointDNN graph model. Finally, we show how the constrained problem is formulated by setting up ILP. 

\subsection{Energy and Latency Profiling}
\label{energy_latency_profiling}
There are three methods in measuring the latency and energy consumption of each layer in neural networks~\cite{energyprofiling}: 

\textbf{Statistical Modeling:} In this method, a regression model over the configurable parameters of operators (e.g. filter size in the convolution) can be used to estimate the associated latency and energy. This method is prone to large errors because of the inter-layer optimizations performed by DNN software packages. Therefore, it is necessary to consider the execution of several consecutive operators grouped during profiling. Many of these software packages are proprietary, making access to inter-layer optimization techniques impossible. 

% NVIDIA\textsuperscript{\textregistered} 

In order to illustrate this issue, we designed two experiments with 25 consecutive convolutions on NVIDIA Pascal\textsuperscript{\texttrademark} GPU using cuDNN\textsuperscript{\textregistered} library~\cite{cuDNN}. In the first experiment, we measure the latency of each convolution operator separately and set the total latency as the sum of them. In the second experiment, we execute the grouped convolutions in a single kernel together and measure the total latency. All parameters are located on the GPU's memory in both experiments, avoiding any data transfer from the main memory to make sure results are exactly representing the actual computation latency. 

% ALLIGN!

\begin{figure}
\centering
\includegraphics{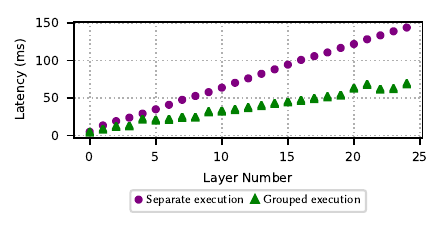}
\caption{Latency of grouped and separated execution of convolution operator.}\label{grouped_execution}
\captionsetup{justification=centering,margin=2cm}
\end{figure}

As we see in Figure~\ref{grouped_execution}, there is a large error gap between separated and grouped execution experiments which grows as the number of convolutions is increased. This observation confirms that we need to profile grouped operators to have more accurate estimations. Considering the various consecutive combination of operators and different input sizes, this method requires a very large number of measurements, not to mention the need for a complex regression model.

% TODO: Erfan: Change it to layer count! Change the explanation! 

\textbf{Analytical Modeling:} To derive analytical formulations for estimating the latency and energy consumption, it is required to obtain the exact hardware and software specifications. However, the state-of-the-art in latency modeling of DNNs~\cite{Paleo} fails to estimate layer-level delay within an acceptable error bound, for instance, underestimating the latency of a fully connected layer with 4096 neurons by around 900\%. Industrial developers do not reveal the detailed hardware architecture specifications and the proprietary parallel computing architectures such as CUDA\textsuperscript{\textregistered}, therefore, the analytical approach could be quite challenging~\cite{GPUMODEL}.

\textbf{Application-specific Profiling:} In this method, the DNN architecture of the application being used is profiled in run-time. The number of applications in a mobile device using neural networks is generally limited. In conclusion, this method is more feasible, promising higher accuracy estimations. We have chosen this method for the estimation of energies and latencies in the experiments of this paper.

\subsection{JointDNN Graph Model}
\begin{figure}[t]
\centering
\includegraphics{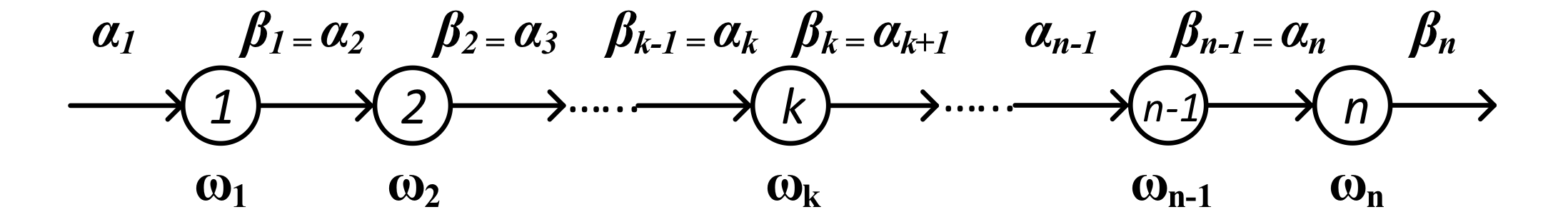}
\caption{Computation model in linear topology.}\label{linear_topology}
\captionsetup{justification=centering,margin=2cm}
\end{figure}

First, we assume that a DNN is presented by a sequence of distinct layers with a linear topology as depicted in Figure~\ref{linear_topology}. Layers are executed sequentially, with output data generated by one layer feeds into the input of the next one. We denote the input and output data sizes of k$^{th}$ layer as $\alpha_k$ and $\beta_k$, respectively. Denoting the latency (energy) of layer k as $\omega_k$, where $k = 1, 2, ..., n$, the total latency (energy) of querying the DNN is $\sum_{k=1}^{n}{\omega_k}$.

% k$^{th}$

\begin{figure}[t]
\centering
\includegraphics{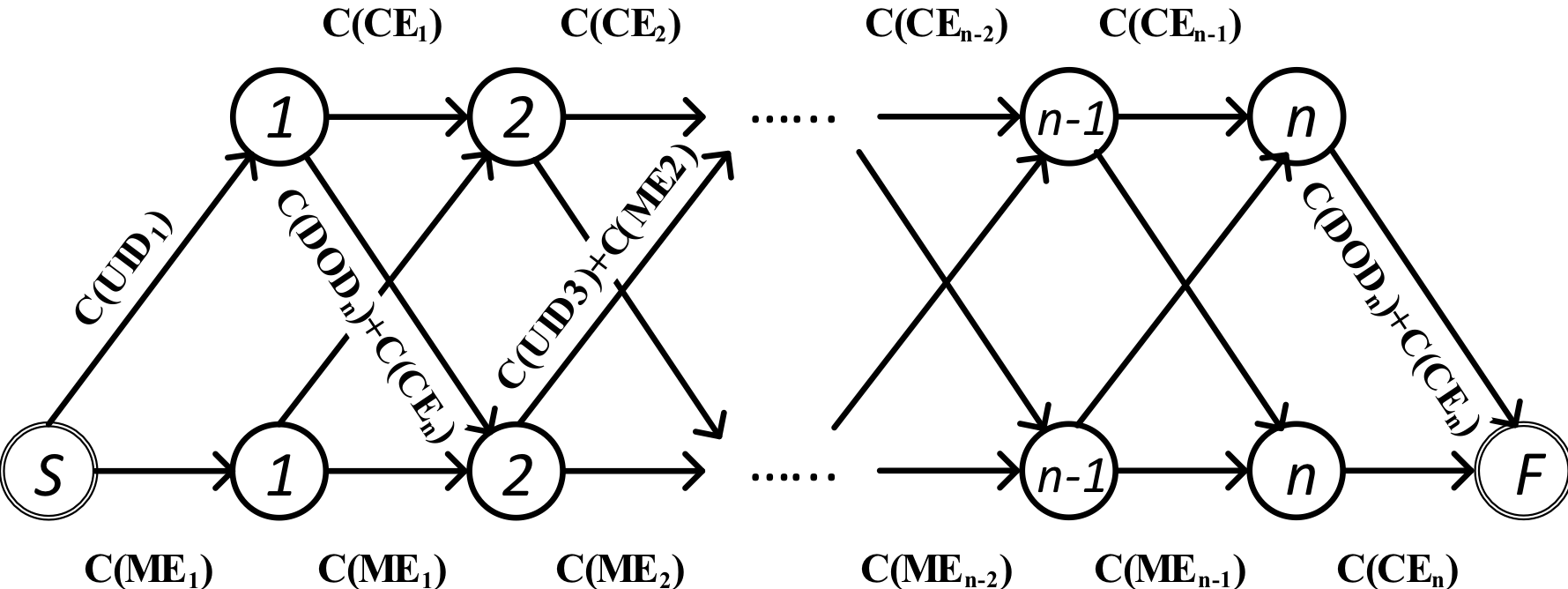}
\caption{Graph representation of mobile cloud computing optimal scheduling problem for linear topology.}
\captionsetup{justification=centering,margin=2cm}
\label{linear_topology_mc}
\end{figure}

\begin{figure*}
\centering
\includegraphics{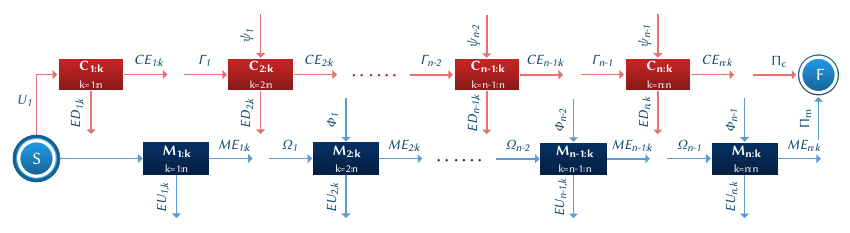}
\captionsetup{justification=centering,margin=2cm}
\caption{JointDNN graph model. The shortest path from S to F determines the schedule of executing the layers on mobile or cloud.}
\label{packing_topology}
\end{figure*}

The mobile cloud computing optimal scheduling problem can be reduced to the shortest path problem, from node $S$ to $F$, in the graph of Figure~\ref{linear_topology_mc}. \textbf{Mobile Execution} cost of the k$^{th}$ layer ($C(ME_k)$) is the cost of executing the k$^{th}$ layer in the mobile while the cloud server is idle. \textbf{Cloud Execution} cost of the k$^{th}$ layer ($C(CE_k)$) is the executing cost of the k$^{th}$ layer in the cloud server while the mobile is idle. \textbf{Uploading the Input Data} cost of the k$^{th}$ layer is the cost of uploading output data of the (k-1)$^{th}$ layer to the cloud server $(UID_k)$. \textbf{Downloading the Input Data} cost of the k$^{th}$ layer is the cost of downloading output data of the (k-1)$^{th}$ layer to the mobile $(DOD_k)$. The costs can refer to either latency or energy. However, as we showed in Section~\ref{energy_latency_profiling}, the assumption of linear topology in DNNs is not true and we need to consider all the consecutive grouping of the layers in the network. This fact suggests the replacement of linear topology by a tournament graph as depicted in Figure~\ref{packing_topology}. We define the parameters of this new graph, \textit{JointDNN graph model}, in Table~\ref{jointDNNGraphParam}. 

\begin{table}[b]
\caption{Parameter Definition of Graph Model} % title of Table
\label{nonlin_inputs} % is used to refer this table in the text
\centering % used for centering table
\begin{tabular}{|c|c|} % centered columns (4 columns)
\hline %inserts double horizontal lines
\textbf{Param.} & \textbf{Description of Cost}\\ [0.5ex] % inserts table
%heading
\hline % inserts single horizontal line
$CE_{i:j}$ & Executing layers $i$ to $j$ on the cloud \\
\hline % inserts single horizontal line
$ME_{i:j}$ & Executing layers $i$ to $j$ on the mobile \\
\hline % inserts single horizontal line
$ED_{i,j}$ & $CE_{i:j}$ + $DOD_j$\\
\hline % inserts single horizontal line
$EU_{i,j}$ & $ME_{i:j}$ + $UID_j$\\
\hline % inserts single horizontal line
$\phi_{k}$ & All the following edges: $\forall i=1:k-1$ $ED_{i,k-1}$\\
\hline % inserts single horizontal line
$\Omega_{k}$ & All the following edges: $\forall i=1:k-1$ $ME_{i,k-1}$\\
\hline % inserts single horizontal line
$\Psi_{k}$ & All the following edges: $\forall i=1:k-1$ $EU_{i,k-1}$\\
\hline % inserts single horizontal line
$\Gamma_{k}$ &  All the following edges: $\forall i=1:k-1$ $CE_{i,k-1}$\\
\hline % inserts single horizontal line
$\Pi_m$ & All the following edges: $\forall i=1:n$ $ME_{i,n}$\\
\hline % inserts single horizontal line
$\Pi_c$ & All the following edges: $\forall i=1:n$ $ED_{i,n}$\\
\hline % inserts single horizontal line
$U_1$ & Uploading the input of the first layer\\
\hline %inserts single line
\end{tabular}
\label{jointDNNGraphParam}
\end{table}

% We introduce two dummy nodes: node S as the starting node and node F as the finishing node. 

% the k$^{th}$ layer

\begin{figure}[b]
\centering
\includegraphics{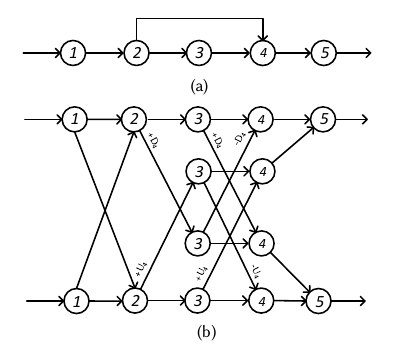}
%\captionsetup{justification=centering,margin=2cm}
\caption{(a) A residual building block in DNNs(b) Transformation of a residual building block to be able to be used in JointDNN's shortest path based scheduler}\label{resnet}
\end{figure}

In this graph, node $C_{i:j}$ represents that the layers $i$ to $j$ are computed on the cloud server, while node $M_{i:j}$ represents that the layers $i$ to $j$ are computed on the mobile device. An edge between two adjacent nodes in JointDNN graph model is associated with four possible cases: 1) A transition from the mobile to the mobile, which only includes the mobile computation cost ($ME_{i,j}$) 2) A transition from the cloud to the cloud, which only includes the cloud computation cost ($CE_{i,j}$) 3) A transition from the mobile to the cloud, which includes the mobile computation cost and uploading cost of the inputs of the next node ($EU_{i,j} = ME_{i,j} + UID_{j+1}$) 4) A transition from the cloud to the mobile, which includes the cloud computation cost and downloading cost of the inputs of the next node ($ED_{i,j} = CE_{i,j} + DOD_{j+1}$). Under this formulation, we can transform the computation scheduling problem to finding the shortest path from $S$ to $F$. Residual networks are a class of powerful and easy-to-train architectures of DNNs~\cite{ResNet}. In residual networks, as depicted in Figure~\ref{resnet} (a), the output of one layer is fed into another layer with a distance of at least two. Thus, we need to keep track of the source layer (node $2$ in Figure~\ref{resnet}) to know that this layer is computed on the mobile or the cloud. Our standard graph model has a memory of one which is the very previous layer. We provide a method to transform the computation graph of this type of network to our standard model, JointDNN graph. 

In this regard, we add two additional chains of size $k-1$, where $k$ is the number of nodes in the residual block ($3$ in Figure~\ref{resnet}). One chain represents the case of computing layer $2$ on the mobile and the other one represents the case of computing layer $2$ on the cloud. In Figure~\ref{resnet}, we have only shown the weights that need to be modified, where $D_2$ and $U_2$ are the cost of downloading and uploading the output of layer $2$, respectively.

By solving the shortest path problem in the JointDNN graph model, we can obtain the optimal scheduling of inference in DNNs. The online training consists of one inference and one back-propagation step. The total number of layers is noted by $N$ consistently throughout this paper so there are $2N$ layers for modeling training, where the second $N$ layers are the mirrored version of the first $N$ layers, and their associated operations are the gradients of the error function concerning the DNN's weights. The main difference between the mobile cloud computing graph of inference and online training is the need for updating the model by downloading the new weights from the cloud. We assume that the cloud server performs the whole back-propagation step separately, even if it is scheduled to be done on the mobile, therefore, there is no need for the mobile device to upload the weights that are updated by itself to save mobile energy consumption. The modification in the JointDNN graph model is adding the costs of downloading weights of the layers that are updated in the cloud to $ED_{i,j}$. The shortest path problem can be solved in polynomial time efficiently. 

However, the problem of the shortest path subjected to constraints is NP-Complete~\cite{NPComplete}. For instance, assuming our standard graph is constructed for energy and we need to find the shortest path subject to the constraint of the total latency of that path is less than a time deadline (QoS). However, there is an approximation solution to this problem, "LARAC" algorithm~\cite{LARAC}, the nature of our application does not require to solve this optimization problem frequently, therefore, we aim to obtain the optimal solution. We can constitute a small look-up table of optimization results for a different set of parameters (e.g. network bandwidth, cloud server load, etc.). We provide the ILP formulations of DNN partitioning in the following sections.

\subsection{ILP Setup}

\subsubsection{Performance Efficient Computation Offloading ILP Setup for Inference}

We formulated the scheduling of inference in DNNs as an ILP with tractable number of variables. In our method, first we profile the delay and energy consumption of consecutive layers of size $m$ $\in$ $\{1, 2,\dots, N\}$. Thus, we will have
  \begin{equation}
      \begin{split}
N + (N-1) + ... + 1 = N(N+1)/2
  \end{split}
\end{equation}
number of different profiling values for delay and energy.
Considering layer $i$ to layer $j$ to be computed either on the mobile device or cloud server, we assign two binary variables $m_{i,j}$ and $c_{i,j}$, respectively. Download and upload communication delays needs to be added to the execution time, when switching from/to cloud to/from mobile, respectively.

\begin{equation}
      \begin{split}
T_{computation} &= \sum_{i=1}^{n}{\sum_{j=i}^{n}{
(m_{i,j}.T_{mobile_{L_{i,j}}} + c_{i,j}.T_{cloud_{L_{i,j}}})
}}
  \end{split}
\end{equation}
  \begin{equation}
      \begin{split}
T_{communication} &= \sum_{i=1}^{n}{\sum_{j=i}^{n}{\sum_{k=j+1}^{n}{m_{i,j}.c_{j+1,k}.T_{upload_{L_j}}}}} \\
& + \sum_{i=1}^{n}{\sum_{j=i}^{n}{\sum_{k=j+1}^{n}{c_{i,j}.m_{j+1,k}.T_{download_{L_j}}}}} \\
& + \sum_{i=1}^{n}{c_{1,i}.T_{upload_{L_i}}} \\
& + \sum_{i=1}^{n}{c_{i,n}.T_{download_{L_n}}}
  \end{split}
  \label{eq: full_pass_comm}
\end{equation}

\begin{equation}
T_{total} = T_{computation} + T_{communication}\thinspace\thinspace\thinspace\thinspace\thinspace\thinspace\thinspace\thinspace\thinspace\thinspace\thinspace\thinspace\thinspace\thinspace\thinspace\thinspace\thinspace\thinspace\thinspace\thinspace\thinspace\thinspace\thinspace\thinspace\thinspace\thinspace\thinspace\thinspace\thinspace\thinspace\thinspace
\end{equation}

$T_{mobile_{L_{i,j}}}$ and $T_{cloud_{L_{i,j}}}$ represent the execution time of the i$^{th}$ layer to the j$^{th}$ layer on the mobile and cloud, respectively. $T_{download_{L_i}}$ and $T_{upload_{L_i}}$ represent the latency of downloading and uploading the output of the i$^{th}$ layer, respectively. Considering each set of the consecutive layers, whenever $m_{i,j}$ and one of $\{c_{j+1,k}\}_{k=j+1:n}$ are equal to one, the output of the j$^{th}$ layer is uploaded to the cloud. The same argument applies to downloading.
We also note that the last two terms in Eq.~\ref{eq: full_pass_comm} represent the condition by which the last layer is computed on the cloud and we need to download the output to the mobile device, and the first layer is computed on the cloud and we need to upload the input to the cloud, respectively. To support for residual architectures, we need to add a pair of download and upload terms similar to the first two terms in Eq.~\ref{eq: full_pass_comm} for the starting and ending layers of each residual block. In order to guarantee that all layers are computed exactly once, we need to add the following set of constraints:

\begin{equation}
	\begin{split}
		\forall m \in {1:n}: \sum_{i=1}^{m}{\sum_{j=m}^{n}{(m_{i,j} + c_{i,j})}} = 1
  	\end{split}
\end{equation}

Because of the non-linearity of multiplication, an additional step is needed to transform Eq.~\ref{eq: full_pass_comm} to the standard form of ILP. We define two sets of new variables:

\begin{equation}
\label{eq: download_upload_binary_var} 
	\begin{aligned}
      &u_{i,j} = m_{i,j}.\sum_{k=j+1}^{n}{c_{j+1,k}} \\
      &d_{i,j} = c_{i,j}.\sum_{k=j+1}^{n}{m_{j+1,k}}
     \end{aligned}
\end{equation}

with the following constraints:

\begin{equation}\label{eq: download_upload_constraints} 
	\begin{aligned}
      &u_{i,j} \leq m_{i,j}\\
      &u_{i,j} \leq \sum_{k=j+1}^{n}{c_{j+1,k}} \\
      &m_{i,j} + \sum_{k=j+1}^{n}{c_{j+1,k}} - u_{i,j} \leq 1\\
      &d_{i,j} \leq c_{i,j}\\
      &d_{i,j} \leq \sum_{k=j+1}^{n}{m_{j+1,k}} \\
      &c_{i,j} + \sum_{k=j+1}^{n}{m_{j+1,k}} - d_{i,j} \leq 1
	\end{aligned}
\end{equation}

The first two constraints ensure that $u_{i,j}$ will be zero if either $m_{i,j}$ or $\sum_{l=j+1}^{n}{c_{j+1,l}}$ are zero. The third inequality guarantees that $u_{i,j}$ will take value one if both binary variables, $m_{i,j}$ and $\sum_{l=j+1}^{n}{c_{j+1,l}}$, are set to one. The same reasoning works for $d_{i,j}$. In summary, the total number of variables in our ILP formulation will be $4N(N+1)/2$, where $N$ is total number of layers in the network. 

\subsubsection{Energy Efficient Computation Offloading ILP Setup for Inference}
Because of the nature of the application, we only care about the energy consumption on the mobile side. We formulate ILP as follows:
  \begin{equation}
      \begin{split}
E_{computation} &= \sum_{i=1}^{n}{\sum_{j=i}^{n}{
m_{i,j}.E_{mobile_{L_{i,j}}}}\thinspace\thinspace\thinspace\thinspace\thinspace\thinspace\thinspace\thinspace\thinspace\thinspace\thinspace\thinspace\thinspace\thinspace\thinspace\thinspace\thinspace\thinspace\thinspace\thinspace\thinspace\thinspace\thinspace\thinspace\thinspace\thinspace\thinspace\thinspace\thinspace\thinspace\thinspace\thinspace\thinspace\thinspace\thinspace\thinspace}\label{eq: energy_comp}
  \end{split}
\end{equation}
  \begin{equation}
      \begin{split}
E_{communication} &= \sum_{i=2}^{n}{\sum_{j=i}^{n}{m_{i,j}.E_{download_{L_i}}}} \\
& + \sum_{i=1}^{n}{\sum_{j=i}^{n-1}{m_{i,j}.E_{upload_{L_j}}}} \\
& + (\sum_{i=1}^{n}{(1-m_{1,i}) - (n-1)).E_{upload_{L_1}}} \\
& + (\sum_{i=1}^{n}{(1-m_{i,n}) - (n-1)).E_{download_{L_n}}}\label{eq: energy_comm}
  \end{split}
\end{equation}
  \begin{equation}
E_{total} = E_{computation} + E_{communication}\thinspace\thinspace\thinspace\thinspace\thinspace\thinspace\thinspace\thinspace\thinspace\thinspace\thinspace\thinspace\thinspace\thinspace\thinspace\thinspace\thinspace\thinspace\thinspace
\end{equation}

$E_{mobile_{L_{i,j}}}$ and $E_{cloud_{L_{i,j}}}$ represent the amount of energy required to compute the i$^{th}$ layer to the j$^{th}$ layer on the mobile and cloud, respectively. $E_{download_{L_i}}$ and $E_{upload_{L_i}}$ represent the energy required to download and upload the output of i$^{th}$ layer, respectively.
Similar to performance efficient ILP constraints, each layer should be executed exactly once:

\begin{equation}
      \begin{split}
\forall m \in {1:n}: \sum_{i=1}^{m}{\sum_{j=m}^{n}{m_{i,j}}} \leq 1
  \end{split}
\end{equation}

The ILP problem can be solved for different set of parameters (e.g. different uplink and download speeds), and then the scheduling results can be stored as a look-up table in the mobile device. Moreover because the number of variables in this setup is tractable solving ILP is quick. For instance, solving ILP for AlexNet takes around 0.045 seconds on Intel(R) Core(TM) i7-3770 CPU with MATLAB\textregistered's intlinprog() function using primal simplex algorithm.\\
\subsubsection{Performance Efficient Computation Offloading ILP Setup for Training}
The ILP formulation of online training phase is very similar to that of inference. In online training we have $2N$ layers instead of $N$ obtained by mirroring the DNN, where the second $N$ layers are backward propagation. Moreover, we need to download the weights that are updated in the cloud to the mobile. We assume that the cloud server always has the most updated version of the weights and does not require the mobile device to upload the updated weights. The following terms need to be added for the ILP setup of training:

\begin{equation}
      \begin{split}
T_{computation} &= \sum_{i=1}^{2n}{\sum_{j=i}^{2n}{
(m_{i,j}.T_{mobile_{L_{i,j}}} + c_{i,j}.T_{cloud_{L_{i,j}}})
}}\label{eq: full_pass_comp}
  \end{split}
\end{equation}
  \begin{equation}
      \begin{split}
T_{communication} &= \sum_{i=1}^{2n}{\sum_{j=i}^{2n}{\sum_{k=j+1}^{2n}{m_{i,j}.c_{j+1,k}.T_{upload_{L_j}}}}} \\
& + \sum_{i=1}^{2n}{\sum_{j=i}^{2n}{\sum_{k=j+1}^{2n}{c_{i,j}.m_{j+1,k}.T_{download_{L_j}}}}} \\
& + \sum_{i=1}^{n}{c_{1,i}.T_{upload_{L_i}}} \\
& + \sum_{i=n+1}^{2n}{\sum_{j=i}^{2n}{c_{i,j}.T_{download_{W_{i}}}}}
  \end{split}
\end{equation}

\begin{equation}
T_{total} = T_{computation} + T_{communication}\thinspace\thinspace\thinspace\thinspace\thinspace\thinspace\thinspace\thinspace\thinspace\thinspace\thinspace\thinspace\thinspace\thinspace\thinspace\thinspace\thinspace\thinspace\thinspace\thinspace\thinspace\thinspace\thinspace\thinspace\thinspace\thinspace\thinspace\thinspace\thinspace\thinspace\thinspace
\end{equation}

\subsubsection{Energy Efficient Computation Offloading ILP Setup for Training}
  \begin{equation}
      \begin{split}
E_{computation} &= \sum_{i=1}^{2n}{\sum_{j=i}^{2n}{
m_{i,j}.E_{mobile_{L_{i,j}}}}\thinspace\thinspace\thinspace\thinspace\thinspace\thinspace\thinspace\thinspace\thinspace\thinspace\thinspace\thinspace\thinspace\thinspace\thinspace\thinspace\thinspace\thinspace\thinspace\thinspace\thinspace\thinspace\thinspace\thinspace\thinspace\thinspace\thinspace\thinspace\thinspace\thinspace\thinspace\thinspace\thinspace\thinspace\thinspace\thinspace}\label{eq: energy_comp_train}
  \end{split}
\end{equation}
  \begin{equation}
      \begin{split}
E_{communication} &= \sum_{i=2}^{2n}{\sum_{j=i}^{2n}{m_{i,j}.E_{download_{L_i}}}} \\
& + \sum_{i=1}^{2n}{\sum_{j=i}^{2n-1}{m_{i,j}.E_{upload_{L_j}}}} \\
& + (\sum_{i=1}^{2n}{(1-m_{1,i}) - (2n-1)).E_{upload_{L_1}}} \\
& + (\sum_{i=n+1}^{2n}\sum_{j=i}^{2n}{(1-m_{i,j}) - (n-1)}).E_{download_{W_{i}}}\label{eq: energy_comm_train}
  \end{split}
\end{equation}
  \begin{equation}
E_{total} = E_{computation} + E_{communication}\thinspace\thinspace\thinspace\thinspace\thinspace\thinspace\thinspace\thinspace\thinspace\thinspace\thinspace\thinspace\thinspace\thinspace\thinspace\thinspace\thinspace\thinspace\thinspace
\end{equation}

\subsubsection{Scenarios}
There can be different optimization scenarios defined for ILP as listed below:
\begin{itemize}
\item \textbf{Performance efficient computation:} In this case, it is sufficient to solve the ILP formulation for performance efficient computation offloading.
\item \textbf{Energy efficient computation:} In this case, it is sufficient to solve the ILP formulation for energy efficient computation offloading.
\item \textbf{Battery budget limitation:} In this case, based on the available battery, the operating system can decide to dedicate a specific amount of energy consumption to each application. By adding the following constraint to the performance efficient ILP formulation, our framework would adapt to battery limitations: 
\begin{equation}
\begin{split}
	E_{computation} + E_{communication} \leq E_{ubound}
\end{split}
\end{equation}

\item\textbf{Cloud limited resources:} In the presence of cloud server congestion or limitations on user's subscription, we can apply execution time constraints to each application to alleviate the server load:

\begin{equation}
\begin{split}
\sum_{i=1}^{n}{\sum_{j=i}^{n}{
c_{i,j}.T_{cloud_{L_{i,j}}}}} \leq T_{ubound}
\end{split}
\end{equation}

\item \textbf{QoS:} In this scenario, we minimize the required energy consumption while meeting a specified deadline:

\begin{equation}
      \begin{split}
min\{E_{computation} + E_{communication}\} \\
T_{computation} + T_{communication} \leq T_{QoS}
  \end{split}
\end{equation}

This constraint could be applied to both energy and performance efficient ILP formulations.
\end{itemize}

\begin{algorithm}
    \SetKwInOut{Input}{Input}
    \SetKwInOut{Output}{Output}

    \underline{function JointDNN} $(N,L_i,D_i,NB,NP)$\;
    \Input{
1: $N$: number of layers in the DNN\\
2: $L_i|i=1:N$: layers in the DNN\\
3: $D_i|i=1:N$: data size at each layer\\
4: $NB$: mobile network bandwidth\\
5: $NP$: mobile network uplink and downlink power consumption\\
}
    \Output{Optimal schedule of DNN}
  \For{$i = 0;\ i < N;\ i = i + 1$}{
  \For{$j = 0;\ j < N;\ j = j + 1$}{
    $Latency_{i,j}, Energy_{i,j}$ = ProfileGroupedLayers$(i,j)$\;
    }
  }
G,S,F = ConstructShortestPathGraph($N$,$L_i$,$D_i$,$NB$,$NP$) //S and F are start and finish nodes and G is the JointDNN graph model\\
    \eIf{no constraints}
      {
          $schedule$ = \textbf{ShortestPath(G,S,F)}
      }
      {
      \If{Battery Limited Constraint} {
    $E_{comm} + E_{comp} \le E_{ubound}$ \\
          $schedule$ = PerformanceEfficientILP($N$,$L_i$,$D_i$,$NB$,$NP$)
      }
            \If{Cloud Server Contraint} {
           $\sum_{i=1}^{n}{\sum_{j=i}^{n}{
c_{i,j}.T_{cloud_{L_{i,j}}}}} \leq T_{ubound}$\\
          $schedule$ = PerformanceEfficientILP($N$,$L_i$,$D_i$,$NB$,$NP$)
      }
            \If{QoS} {
            $T_{comm} + T_{comp} \le T_{QoS}$ \\
          $schedule$ = EnergyEfficientILP($N$,$L_i$,$D_i$,$NB$,$NP$)
      };\

      }
       return $schedule$\;
    \caption{JointDNN engine optimal scheduling of DNNs}
\end{algorithm}

\section{Evaluation}

\subsection{Deep Architecture Benchmarks}

Since the architecture of neural networks depends on the type of application, we have chosen three common application types of DNNs as shown in Table~\ref{benchmarks}:

\begin{enumerate}
\item \textbf{Discriminative neural networks} are a class of models in machine learning for modeling the conditional probability distribution $P(y|x)$. This class generally is used in classification and regression tasks. AlexNet\cite{AlexNet}, OverFeat\cite{OverFeat}, VGG16\cite{VGG16}, Deep Speech\cite{DeepSpeech}, ResNet\cite{ResNet}, and NiN\cite{NiN} are well-known discriminative models we use as benchmarks in this experiment. Except for Deep Speech, used for speech recognition, all other benchmarks are used in image classification tasks. 

\item \textbf{Generative neural networks} model the joint probability distribution $P(x,y)$, allowing generation of new samples. These networks have applications in Computer Vision~\cite{GAN} and Robotics~\cite{DeepVisual}, which can be deployed on a mobile device. Chair~\cite{Chair} is a generative model we use as a benchmark in this work.
\item \textbf{Autoencoders} are another class of neural networks used to learn a representation for a data set. Their applications are image reconstruction, image to image translation, and denoising to name a few. Mobile robots can be equipped with autoencoders to be used in their computer vision tasks. We use Pix2Pix~\cite{Pix2Pix}, as a benchmark from this class.
\end{enumerate}

\begin{table}
\caption{Benchmark Specifications}
\centering
\begin{adjustbox}{center}
     %\begin{adjustbox}{center}
        \begin{tabular}{|c|c|c|}
        \hline
        \textbf Type & \textbf Model & \textbf Layers \\                
            \hline
            \multirow{4}{*}{Discriminative} & \multicolumn{1}{c}{AlexNet} & \multicolumn{1}{|c|}{21} \\\cline{2-3}
            & \multicolumn{1}{c}{OverFeat} & \multicolumn{1}{|c|}{14} \\\cline{2-3}
            & \multicolumn{1}{c}{Deep Speech} & \multicolumn{1}{|c|}{10} \\\cline{2-3}
            & \multicolumn{1}{c}{ResNet} & \multicolumn{1}{|c|}{70} \\\cline{2-3}
            & \multicolumn{1}{c}{VGG16} & \multicolumn{1}{|c|}{37} \\\cline{2-3}
            & \multicolumn{1}{c}{NiN} & \multicolumn{1}{|c|}{29} \\\hline
           Generative & Chair & 10 \\                \hline
            Autoencoder &  Pix2Pix & 32 \\                \hline
        \end{tabular}
     \end{adjustbox}
%     \vspace{ - 05 mm}

\label{benchmarks}
\end{table}

\begin{table}[h]
    \caption{Mobile networks specifications in the U.S.} 
    \centering 
    \begin{tabular}{|c|c|c|c|} \hline
    \textbf{Param.} & \textbf{3G} & \textbf{4G} & \textbf{Wi-Fi} \\ \hline
    Download speed (Mbps)    &     2.0275     &    13.76     &    54.97 \\ \hline
    Upload speed (Mbps)    &    1.1        &    5.85    &    18.88 \\ \hline
$\alpha_u$ (mW/Mbps)    & 868.98    & 438.39    &    283.17    \\ \hline
$\alpha_d$ (mW/Mbps)     & 122.12     & 51.97    &    137.01        \\ \hline
$\beta$ (mW)    & 817.88    & 1288.04    &    132.86            \\ \hline
\end{tabular}
\label{table:network_parameters} 
\end{table}

\subsection{Mobile and Server Setup}
We used the Jetson TX2 module developed by NVIDIA\textsuperscript{\textregistered}~\cite{JetsonTX2}, a fair representation of mobile computation power as our mobile device. This module enables efficient implementation of DNN applications used in products such as robots, drones, and smart cameras. It is equipped with NVIDIA Pascal\textregistered GPU with 256 CUDA cores and a shared 8~GB 128~bit LPDDR4 memory between GPU and CPU. To measure the power consumption of the mobile platform, we used INA226 power sensor~\cite{INA226}. 

NVIDIA\textsuperscript{\textregistered} Tesla\textsuperscript{\textregistered} K40C~\cite{TeslaGPU} with 12~GB memory serves as our server GPU. The computation capability of this device is more than one order of magnitude compared to our mobile device. 
% TODO: Should I mention that we are not measuring the server power?
% TODO: Should we make the rows of the two tables the same? System? 
% TODO: textregistered sign in table 
% Reference on the table title?

\subsection{Communication Parameters}
To model the communication between platforms, we used the average download and upload speed of mobile Internet~\cite{MobNet, Speedtest} for different networks (3G, 4G and Wi-Fi) as shown in Table~\ref{table:network_parameters}.

The communication power for download ($P_d$) and upload ($P_u$) is dependent on the network throughput ($t_d$ and $t_u$). Comprehensive examinations in~\cite{4GLTE} indicates that uplink and downlink power can be modeled with linear equations (Eq.~\ref{eq:power_model_eq}) fairly accurate with less than 6\% error rate. Table~\ref{table:network_parameters} shows the parameter values of this equation for different networks. 

\begin{equation}
\begin{split}
P_u = \alpha_u t_u + \beta \\
P_d = \alpha_d t_d + \beta \label{eq:power_model_eq}
  \end{split}
\end{equation}

\section{Results} \label{Results}
\begin{figure*}
\centering
\includegraphics{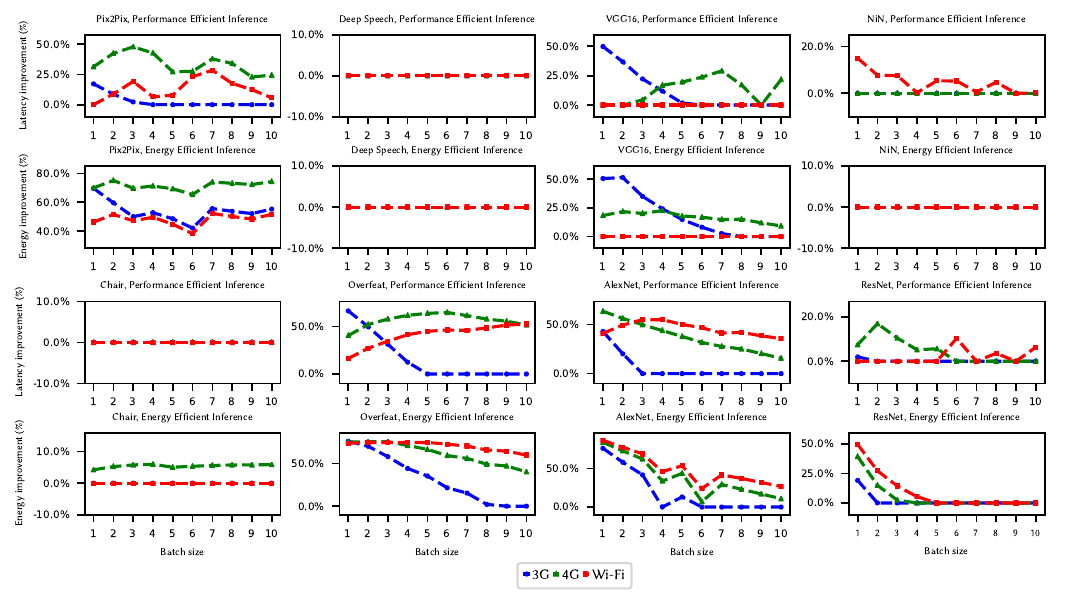}
\captionsetup{justification=centering}
\caption{Latency and energy improvements for different batch sizes during inference over the base case of mobile-only and cloud-only approaches.}
\label{inference_results}
\end{figure*}

\begin{figure*}
\centering
\includegraphics[width=\linewidth]{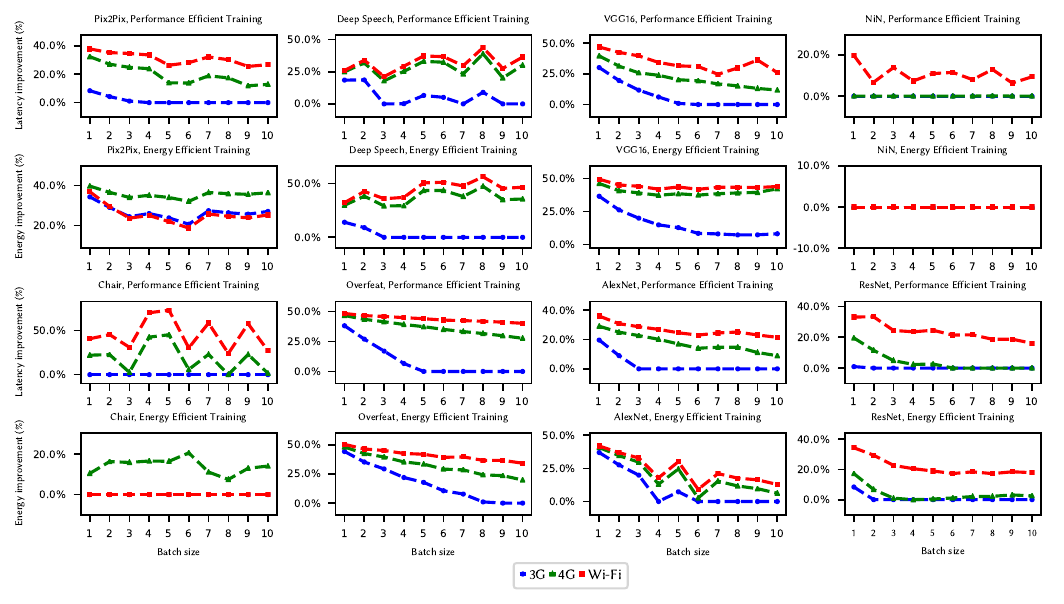}
\captionsetup{justification=centering}
\caption{Latency and energy improvements for different batch sizes during training over the base case of mobile-only and cloud-only approaches.}
\label{training_results}
\end{figure*}
\begin{figure}[b]
\centering
\includegraphics[width=\linewidth]{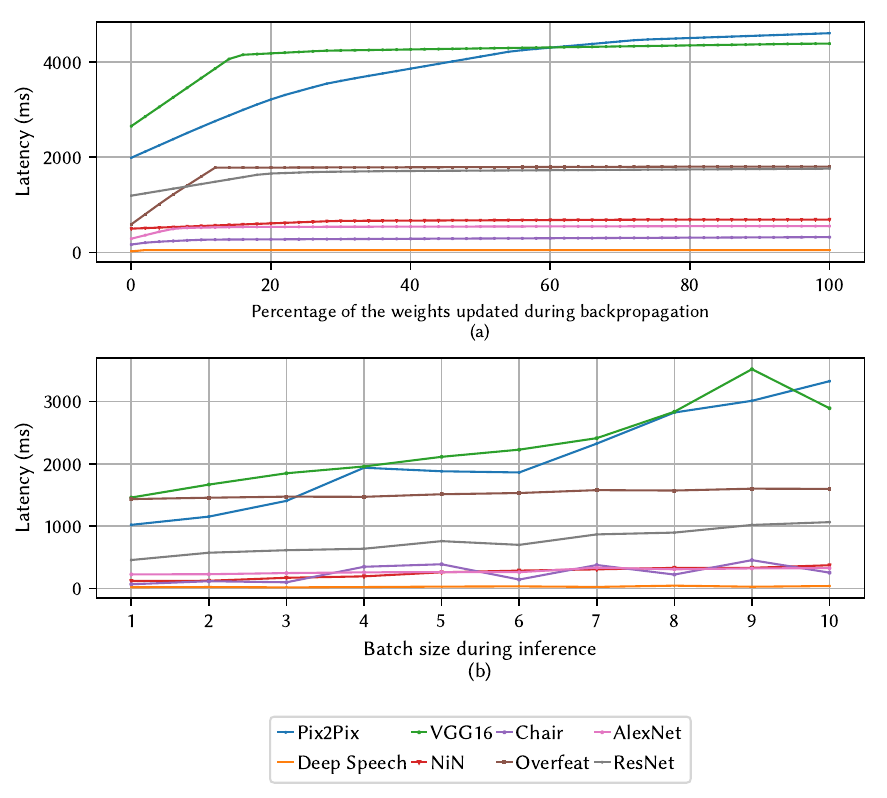}
\caption{(a)~Latency of one epoch of online training using JointDNN algorithm vs percentage of updated weights (b)~Latency of mobile-only inference vs. batch size.}\label{weights_changed}
\end{figure}

The latency and energy improvements of inference and online training with our engine for 8 different benchmarks are shown in Figures~\ref{inference_results} and~\ref{training_results},  respectively. We considered the best case of mobile-only and cloud-only as our baseline. JointDNN can achieve up to 66\% and 86\% improvements in latency and energy consumption, respectively during inference. Communication cost increases linearly with batch size while this is not the case for computation cost and it grows with a much lower rate, as depicted in~\ref{weights_changed}(b). Therefore, a key observation is that as we increase the batch size, the mobile-only approach becomes more preferable.

During online training, the huge communication overhead of transmitting the updated weights will be added to the total cost. Therefore, to avoid downloading this large data, only a few back-propagation steps are computed in the cloud server. We performed a simulation by varying the percentage of updated weight. As the percentage of updated weights increases, the latency and energy consumption becomes constant which is shown in~Figure~\ref{weights_changed}. This is the result of the fact that all the backpropagations will be performed on the mobile device and weights are not transferred from the cloud to the mobile. JointDNN can achieve improvements up to 73\% in latency and 56\% in energy consumption during inference.

\begin{figure}[t]
\centering
\includegraphics{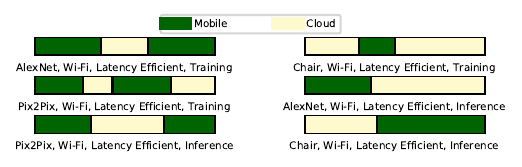}
\caption{Interesting schedules of execution for three types of DNN architectures while mobile/cloud are allowed to use up to half of their computing resources.} \label{interesting_patterns}
\end{figure}

Different patterns of scheduling are demonstrated in Figure~\ref{interesting_patterns}. They represent the optimal solution in the Wi-Fi network while optimizing for latency while mobile/cloud is allowed to use up to half of their computing resources. They show how the computations in DNN is divided between the mobile and the cloud. As can be seen, discriminative models (e.g. AlexNet), inference follows a mobile-cloud pattern and training follows a mobile-cloud-mobile pattern. The intuition is that the last layers are computationally intensive (fully connected layers) but with small data sizes, which require a low communication cost, therefore, the last layers tend to be computed on the cloud. For generative models (e.g. Chair), the execution schedule of inference is the opposite of discriminative networks, in which the last layers are generally huge and in the optimal solution they are computed on the mobile. The reason behind not having any improvement over the base case of mobile-only is that the amount of transferred data is large. Besides, cloud-only becomes the best solution when the amount of transferred data is small (e.g. generative models). Lastly, for autoencoders, where both the input and output data sizes are large, the first and last layers are computed on the mobile.

% \begin{figure}
% \includegraphics{cloud_improvement}
% \caption{Workload reduction of the cloud server in different mobile networks} \label{cloud_improvement}
% \end{figure}

JointDNN pushes some parts of the computations toward the mobile device. As a result, this will lead to less workload on the cloud server. As we see in Table~\ref{cloud_improvement}, we can reduce the cloud server's workload up to 84\% and 53\% on average, which enables the cloud provider to provide service to more users, while obtaining higher performance and lower energy consumption compared to single-platform approaches.  

\begin{table}[h]
\centering
\caption{Workload reduction of the cloud server in different mobile networks}
\label{cloud_improvement}
\begin{tabular}{|c|c|c|c|}
\hline
\textbf{Optimization Target} & \textbf{3G (\%)} & \textbf{4G (\%)} & \textbf{Wi-Fi (\%)} \\ \hline
Latency                      & 84               & 49               & 12                  \\ \hline
Energy                       & 73               & 49               & 51                  \\ \hline
\end{tabular}
\end{table}

\subsection{Communication Dominance}
\begin{figure}[t]
\centering
\includegraphics[width=\linewidth]{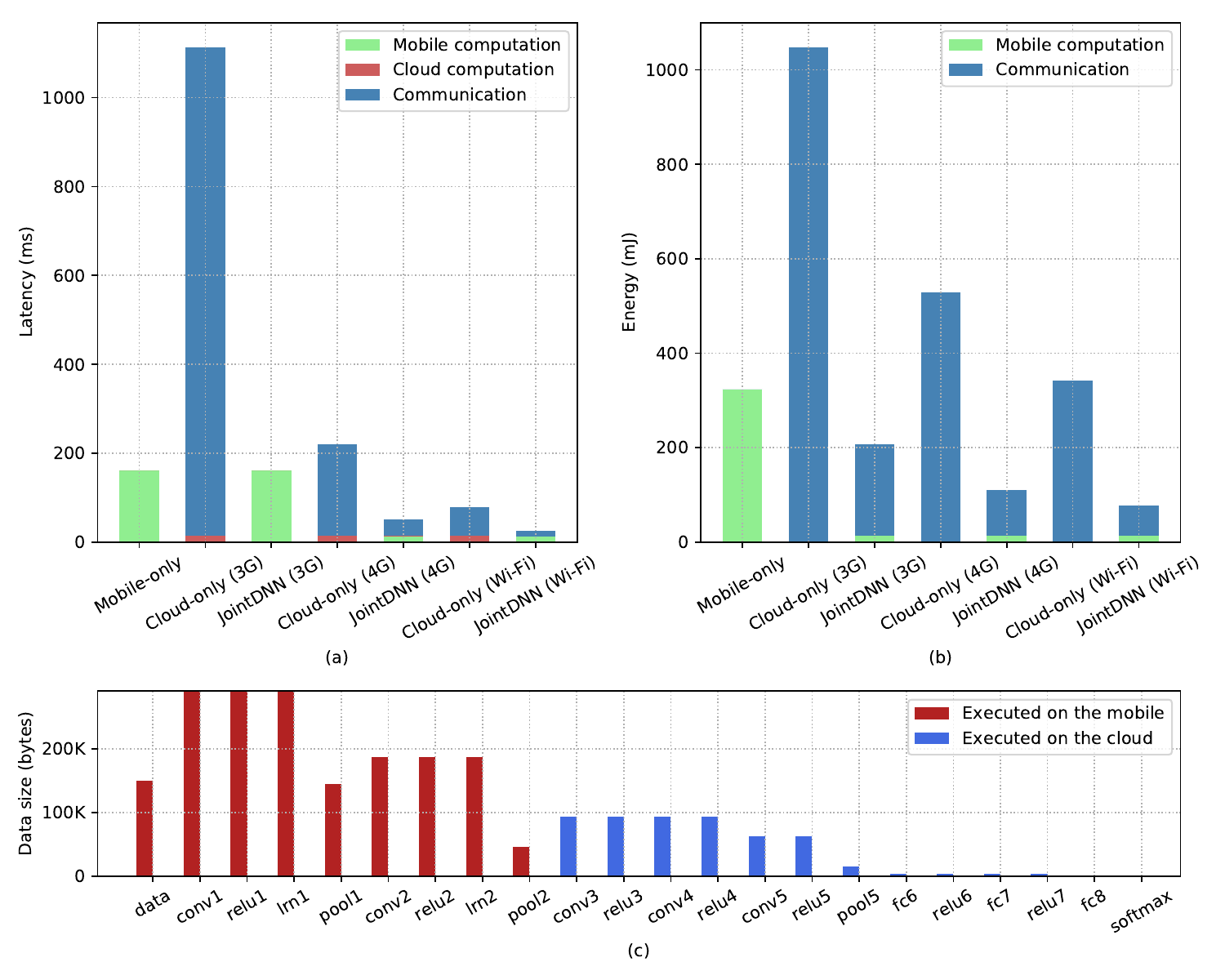}
\caption{(a) Execution time of AlexNet optimized for performance (b) Mobile energy consumption of AlexNet optimized for energy (c) Data size of the layers in AlexNet and the scheduled computation, where the first nine layers are computed on the mobile and the rest on the cloud, which is the optimal solution w.r.t. both performance and energy.}
\label{alexnet_extracted}
\end{figure}

Execution time and energy breakdown for AlexNet, which is noted as a representative for the state-of-the-art architectures deployed in cloud servers, is depicted in Figure~\ref{alexnet_extracted}. The cloud-only approach is dominated by the communication costs. As demonstrated in Figure~\ref{alexnet_extracted}, 99\%, 93\% and 81\% of the total execution time are used for communication in case of 3G, 4G, and Wi-Fi, respectively. This relative portion also applies to energy consumption. Comparing the latency and energy of the communication to those of mobile-only approach, we notice that the mobile-only approach for AlexNet is better than the cloud-only approach in all the mobile networks. We apply loss-less compression methods to reduce the overheads of communication, which will be covered in the next section. 

\subsection{Layer Compression}

The preliminary results of our experiments show that more than $75\%$ of the total energy and delay cost in DNNs are caused by communication in the collaborative approach. This cost is directly proportional to the size of the layer being downloaded to or uploaded from the mobile device. Because of the complex feature extraction process of DNNs, the size of some of the intermediate layers are even larger than the network's input data. For example, this ratio can go as high as $10\times$ in VGG16. To address this bottleneck, we investigated the compression of the feature data before any communication. This process can be applied to different DNN architecture types; however, we only considered CNNs due to their specific characteristics explained later in detail. 

\begin{figure}[b]
\centering
\includegraphics[width=\linewidth]{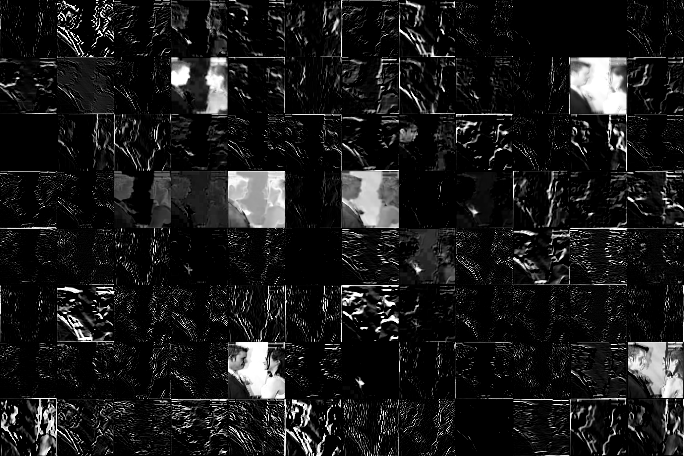}
\caption{Layer output after passing the input image through convolution, normalization and ReLU~\cite{ReLUpaper} layers. Channels are preserving the general structure of the input image and large ratio of the output data is black (zero) due to existence of \textit{relu}. Tiling is used to put all 96 channels together.}
\label{fig:picCNN}
\end{figure}

CNN architectures are mostly used for image and video recognition applications. Because of the spatially local preservation characteristics of \textit{conv} layers, we can assume that the outputs of the first convolution layers are following the same structure as the input image, as shown in Figure~\ref{fig:picCNN}. Moreover, a big ratio of layer outputs is expected to be zero due to the presence of the ReLU layer. Our observations shows that the ratio of neurons equal to zero \textit{(ZR)} varies from 50\% to 90\% after \textit{relu} in CNNs. These two characteristics, layers being similar to the input image, and a large proportion of their data being a single value, suggest that we can employ existing image compression techniques to their output.

% This can be ignored! Especially if we don't want to bring that one PNG and JPEG sentence! 
There are two general categories of compression techniques, lossy and loss-less~\cite{InformationTheoryCover}. In loss-less techniques, the exact original information is reconstructed. On the contrary, lossy techniques use approximations and the original data cannot be reconstructed. In our experiments, we examined the impact of compression of layer outputs using PNG, a loss-less technique, based on the encoding of frequent sequences in an image. 

% Quantization %% UP TO HERE!
Even though the data type of DNN parameters in typical implementations is 32-bits floating-points, most image formats are based on 3-bytes RGB color triples. Therefore, to compress the layer in the same way as 2D pictures, the floating-point data should be quantized into 8-bits fixed-point. Recent studies show representing the parameters of DNNs with only 4-bits affects the accuracy, not more than 1\%~\cite{efficientDNN}. In this work, we implemented our architectures with an 8-bits fixed-point and presented our baseline without any compression and quantization. 
% Quantization can be applied simply by linear mapping considering the uniform distance between each quantization level and converted the 32-bits floating-point into 8-bits. 
% loss-less compression PNG
The layers of CNN contain numerous channels of 2D matrices, each similar to an image. A simple method is to compress each channel separately. In addition to extra overhead of file header for each channel, this method will not take the best of the frequent sequence decoding of PNG. One alternative is locating different channels side by side, referred to as tiling, to form a large 2D matrix representing one layer as shown in Figure~\ref{fig:picCNN}. It should be noted that 1D fully connected layers are very small and we did not apply compression on them.

\begin{figure}[h]
\centering
\includegraphics[width=\linewidth]{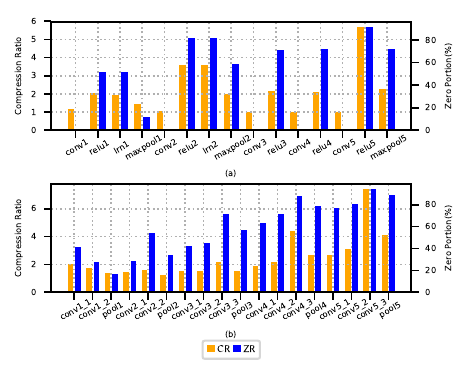}
\caption{Compression Ratio (CR) and ratio of zero valued neurons (ZR) for different layers of (a) AlexNet and (b) VGG16.}
\label{fig:picCR_VGG}
\end{figure}

The Compression Ratio \textit{(CR)} is defined as the ratio of the size of the layer (8-bit) to the size of the compressed 2D matrix in PNG. Looking at the results of compression for two different CNN architectures in Figure~\ref{fig:picCR_VGG}, we can observe a high correlation between the ratio of pixels being zero \textit{(ZR)} and \textit{CR}. PNG can compress the layer data up to $5.8\times$ and $3.5\times$ by average, therefore the communication costs can be reduced drastically. By replacing the compressed layer's output and adding the cost of the compression process itself, which is negligible compared to DNN operators, in JointDNN formulations, we achieve an extra $4.9\times$ and $4.6\times$ improvements in energy and latency on average, respectively.

\section{Related work and comparison}

\textbf{General Task Offloading Frameworks.} There are existing prior arts focusing on offloading computation from the mobile to the cloud\cite{Oedessa, Comet, CloneCloud, MAUI, ExecutionJavaScript, Refactoring, Kumar2013}. However, all these frameworks share a limiting feature that makes them impractical for computation partitioning of the DNN applications. 

These frameworks are programmer annotations dependent as they make decisions about pre-specified functions, whereas JointDNN makes scheduling decisions based on the model topology and mobile network specifications in run-time. Offloading in function level, cannot lead to efficient partition decisions due to layers of a given type within one architecture can have significantly different computation and data characteristics. For instance, a specific convolution layer structure can be computed on mobile or cloud in different models in the optimal solution. 

Neurosurgeon~\cite{Neurosurgeon} is the only prior art exploring a similar computation offloading idea in DNNs between the mobile device and the cloud server at layer granularity. Neurosurgeon assumes that there is only one data transfer point and the execution schedule of the efficient solution starts with mobile and then switches to the cloud, which performs the whole rest of the computations. Our results show this is not true especially for online training, where the optimal schedule of execution often follows the mobile-cloud-mobile pattern. Moreover, generative and autoencoder models follow a multi-transfer points pattern. Also, the execution schedule can start with the cloud especially in case of generative models where the input data size is large. Furthermore, inter-layer optimizations performed by DNN libraries are not considered in Neurosurgeon. Moreover, Neurosurgeon only schedules for optimal latency and energy, while JointDNN adapts to different scenarios including battery limitation, cloud server congestion, and QoS. Lastly, Neurosurgeon only targets simple CNN and ANN models, while JointDNN utilizes a graph-based approach to handle more complex DNN architectures like ResNet and RNNs.

\section{Acknowledgements}
This research was supported by grants from NSF SHF, DARPA MTO, and USC Annenberg Fellowship.
\section{Conclusions and Future Work}
In this paper, we demonstrated that the status-quo approaches, cloud-only or mobile-only, are not optimal with regard to latency and energy. We reduced the problem of partitioning the computations in a DNN to shortest path problem in a graph. Adding constraints to the shortest path problem makes it NP-Complete, therefore, we also provided ILP formulations to cover different possible scenarios of limitations of mobile battery, cloud congestion, and QoS. The output data size in discriminative models is typically smaller than other layers in the network, therefore, last layers are expected to be computed on the cloud, while first layers are expected to be computed on the mobile. Reverse reasoning works for Generative models. Autoencoders have large input and output data sizes, which implies that the first and last layers are expected to be computed on the mobile. With these insights, the execution schedule of DNNs can possibly have various patterns depending on the model architecture in model cloud computing. JointDNN formulations are designed for feed-forward networks and its extension to recurrent neural networks will be studied as a future work.

\ifCLASSOPTIONcaptionsoff
  \newpage
\fi

\bibliographystyle{IEEEtran}
\bibliography{sample-bibliography}
\begin{IEEEbiography}[{\includegraphics[width=1in,height=1in,clip,keepaspectratio]{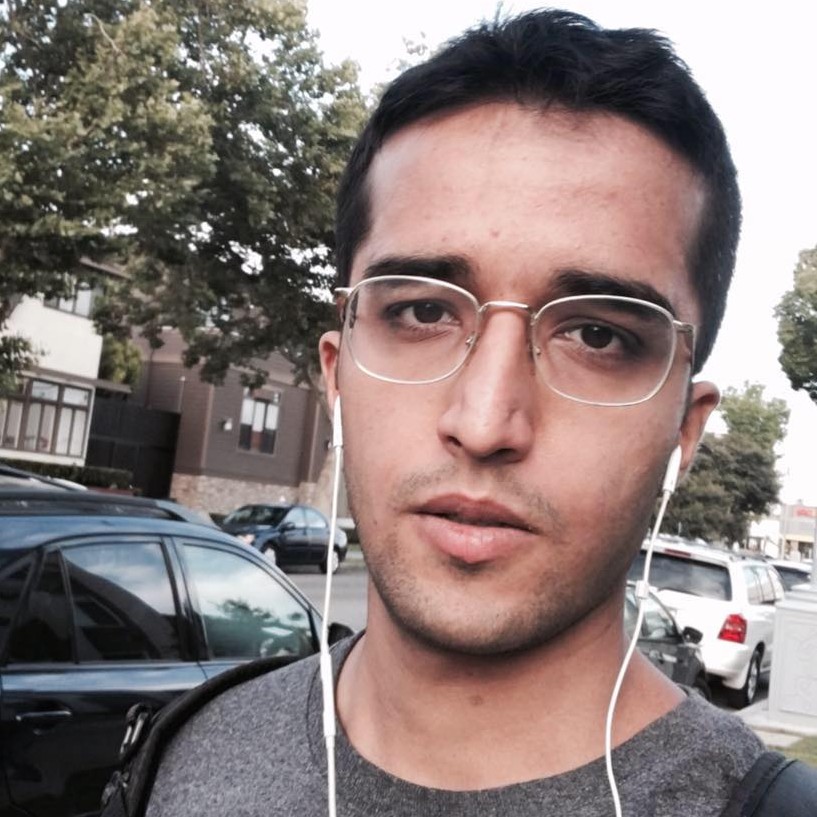}}]{Amir Erfan Eshratifar}
Amir Erfan Eshratifar received dual B.S. degrees in Electrical Engineering and Computer Science from Sharif University of Technology, Tehran, Iran in 2017. He is currently a Ph.D. student in Ming Hsieh Department of Electrical Engineering, University of Southern California (USC), Los Angeles, CA, USA since 2017.
\end{IEEEbiography}

\begin{IEEEbiography}[{\includegraphics[width=1in,height=1in,clip,keepaspectratio]{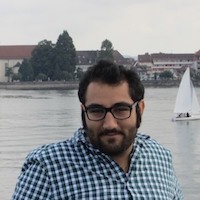}}]{Mohammad Saeed Abrishami}
Mohammad Saeed Abrishami received his B.S. degree in Electrical Engineering from University of Tehran, Tehran, Iran in 2014. He is currently a Ph.D. student in Ming Hsieh Department
of Electrical Engineering, University of Southern California (USC), Los Angeles, CA, USA since 2014.
\end{IEEEbiography}

% insert where needed to balance the two columns on the last page with
% biographies
%\newpage

\begin{IEEEbiography}[{\includegraphics[width=1in,height=1in,clip,keepaspectratio]{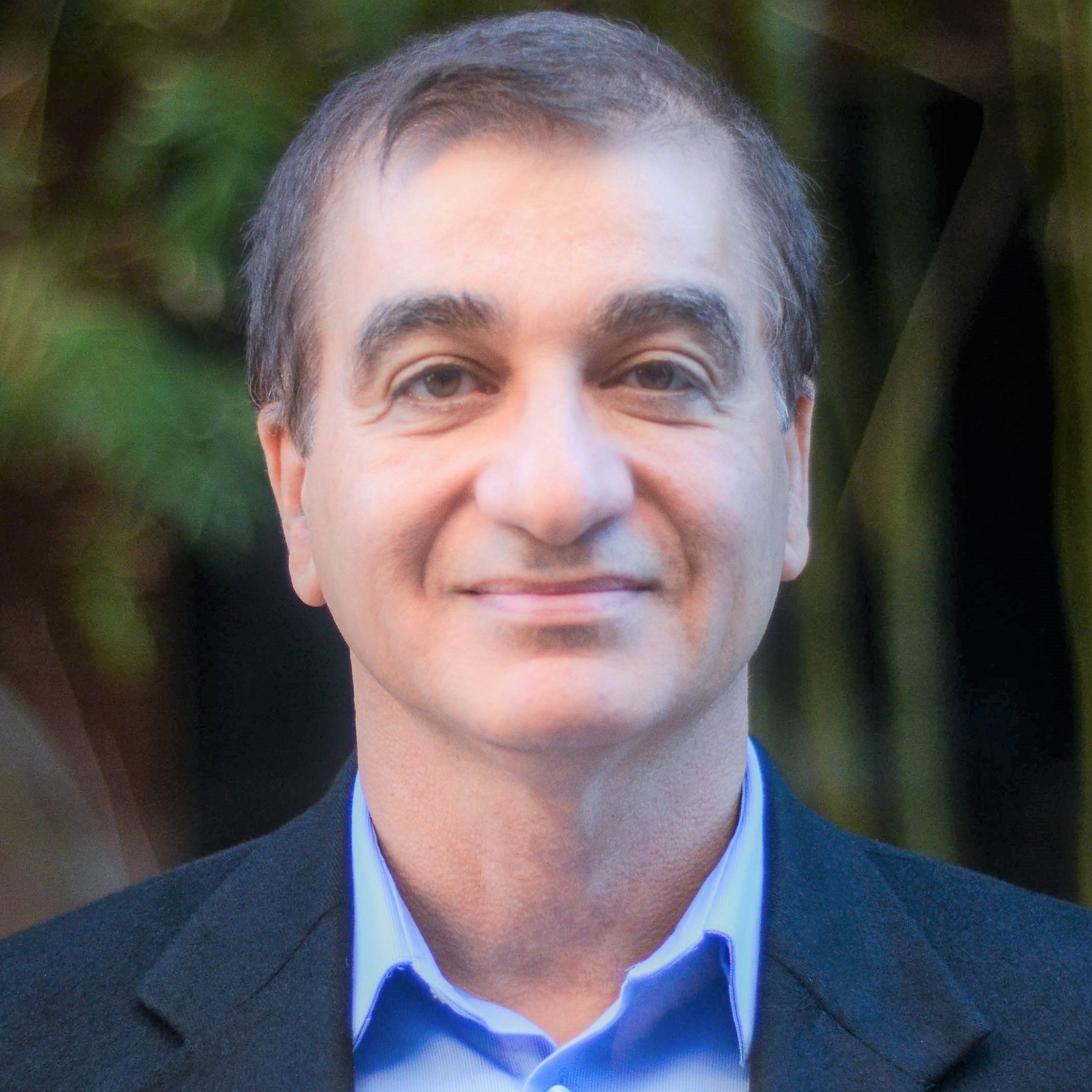}}]{Massoud Pedram}
Massoud Pedram (F’01) received the B.S. degree
in electrical engineering from the California Institute
of Technology, Pasadena, CA, USA, in 1986, and the
M.S. and Ph.D. degrees in electrical engineering and
computer sciences from the University of California Berkeley, CA, USA, in 1989 and 1991, respectively. In 1991, he joined the Ming Hsieh Department
of Electrical Engineering, University of Southern California (USC), Los Angeles, CA, USA, where he is currently the Charles Lee Powel Professor
of USC Viterbi School of Engineering.
\end{IEEEbiography}

% You can push biographies down or up by placing
% a \vfill before or after them. The appropriate
% use of \vfill depends on what kind of text is
% on the last page and whether or not the columns
% are being equalized.

%\vfill

% Can be used to pull up biographies so that the bottom of the last one
% is flush with the other column.
%\enlargethispage{-5in}

% that's all folks
\end{document}